\documentclass[%
 reprint,
superscriptaddress,
 amsmath,amssymb,
 aps,
]{revtex4-1}

\usepackage{graphicx}% Include figure files
\usepackage{dcolumn}% Align table columns on decimal point
\usepackage{bm}% bold math
\begin{document}

%\preprint{APS/123-QED}

\title{Highly charged droplets of superfluid helium}% Force line breaks with \\

\author{Felix Laimer}
\author{Lorenz Kranabetter}%
\author{Lukas Tiefenthaler}
\author{Simon Albertini}
\affiliation{%
 Institut f\"{u}r Ionenphysik und Angewandte Physik, Universit\"{a}t Innsbruck, Technikerstr.~25, A-6020 Innsbruck, Austria}%

\author{Fabio Zappa}
\affiliation{%
 Institut f\"{u}r Ionenphysik und Angewandte Physik, Universit\"{a}t Innsbruck, Technikerstr.~25, A-6020 Innsbruck, Austria}%
\affiliation{%
 Departamento de F\'{i}sica-ICE, Universidade Federal de Juiz de Fora, Campus Universit\'{a}rio, 36036-900, Juiz de Fora, MG, Brazil}%

\author{Andrew M.\ Ellis}
\affiliation{%
 Department of Chemistry, University of Leicester, University Road, Leicester, LE1 7RH, United Kingdom}%

\author{Michael Gatchell}
  \email{michael.gatchell@uibk.ac.at}
\affiliation{%
 Institut f\"{u}r Ionenphysik und Angewandte Physik, Universit\"{a}t Innsbruck, Technikerstr.~25, A-6020 Innsbruck, Austria}%

\affiliation{
 Department of Physics, Stockholm University, 106 91 Stockholm, Sweden}%

\author{Paul Scheier}
\affiliation{%
 Institut f\"{u}r Ionenphysik und Angewandte Physik, Universit\"{a}t Innsbruck, Technikerstr.~25, A-6020 Innsbruck, Austria}%

\date{\today}% It is always \today, today,

\begin{abstract}

We report on the production and study of stable, highly charged droplets of superfluid helium. Using a novel experimental setup we produce neutral beams of liquid helium nanodroplets containing millions of atoms or more that can be ionized by electron impact, mass-per-charge selected, and ionized a second time before being analyzed. Droplets containing up to 55 net positive charges are identified and the appearance sizes of multiply charge droplets are determined as a function of charge state. We show that the droplets are stable on the millisecond time scale of the experiment and decay through the loss of small charged clusters, not through symmetric Coulomb explosions. 

\end{abstract}

\maketitle

%\tableofcontents

%\section{Introduction}
Since their introduction several decades ago, liquid helium nanodroplets have been used to study a wide range of unusual physical and chemical phenomena \cite{Northby:2001aa,Toennies:2004aa,Stienkemeier:2006aa,Mauracher:2018aa}. These droplets allow the investigation of superfluid behavior on the nanoscale, often through probing of the weak interaction of the helium with a dopant molecule located within the droplet \cite{Toennies:2004aa,Stienkemeier:2006aa,Mauracher:2018aa}. Alternatively, this weak interaction with helium can be exploited in spectroscopic studies of atoms, molecules and their clusters \cite{Stienkemeier:2006aa,Campbell:2015aa,Kuhn:2016aa}. Recently, experiments have been performed utilizing new ultrafast diffraction technology to establish the sizes and shapes of individual helium nanodroplets \cite{Gomez:2014aa}. However, the ionization of helium nanodroplets has long thought to be a largely settled matter, with most studies showing singly charged cluster ions emanating from droplets subjected to electron ionization \cite{Mauracher:2018aa}. The possibility of creating multiply charged helium droplets has rarely been considered and there is no prior evidence for species other than doubly charged droplets \cite{Farnik:1997bd}. Using a new experimental approach, this study shows that helium droplets with at least several tens of charges are readily formed at sufficiently high electron energies and electron currents. Furthermore, these ions are stable on the millisecond timescale of these experiments. Evidence is presented that the charges are distributed as multiple single charge sites across the droplets which are kept apart by Coulomb repulsion. These multiply charged helium droplets offer the potential for other new and transformatory experiments, including for the nucleation of clusters and nanoparticles and as a new means of molecular ion spectroscopy based on helium tagging.

Neutral He droplets are formed in the expansion of He gas (Messer, 99.9999\% purity) with stagnation pressure of 20--25 bar through a 5\,$\mu$m nozzle orifice in a copper block that is mounted to the second stage of a closed circuit cryocooler. The temperature of the nozzle can be controlled down to 4.2\,K and depending on the temperature, different mechanisms will dictate the size distribution of the droplets that are formed \cite{Toennies:2004aa}. The droplets pass through a skimmer on their way into the first ionization source where they are ionized by the impact of electrons with kinetic energies up to a few hundred eV. Charged droplets are then mass-per-charge-selected by a spherical electrostatic analyzer. The $m/z$-selected charged droplets can then be ionized further by a second electron impact ionization source. A second electrostatic analyzer, identical to the first one, is then employed to analyze the final mass per charge ratio of the droplets, which are detected with a single channel electron multiplier detector. The velocity spread of droplets in the beam from a continuously operated nozzle is exceptionally small and the average velocity depends strongly on the temperature of the helium before the expansion. This is used to determine the absolute sizes of our droplets \cite{Henne:1996aa,Henne:1998aa}. More experimental details are given in the supporting information.

\begin{figure*}[] %  figure placement: here, top, bottom, or page
   \centering
   \includegraphics[width=5.5in]{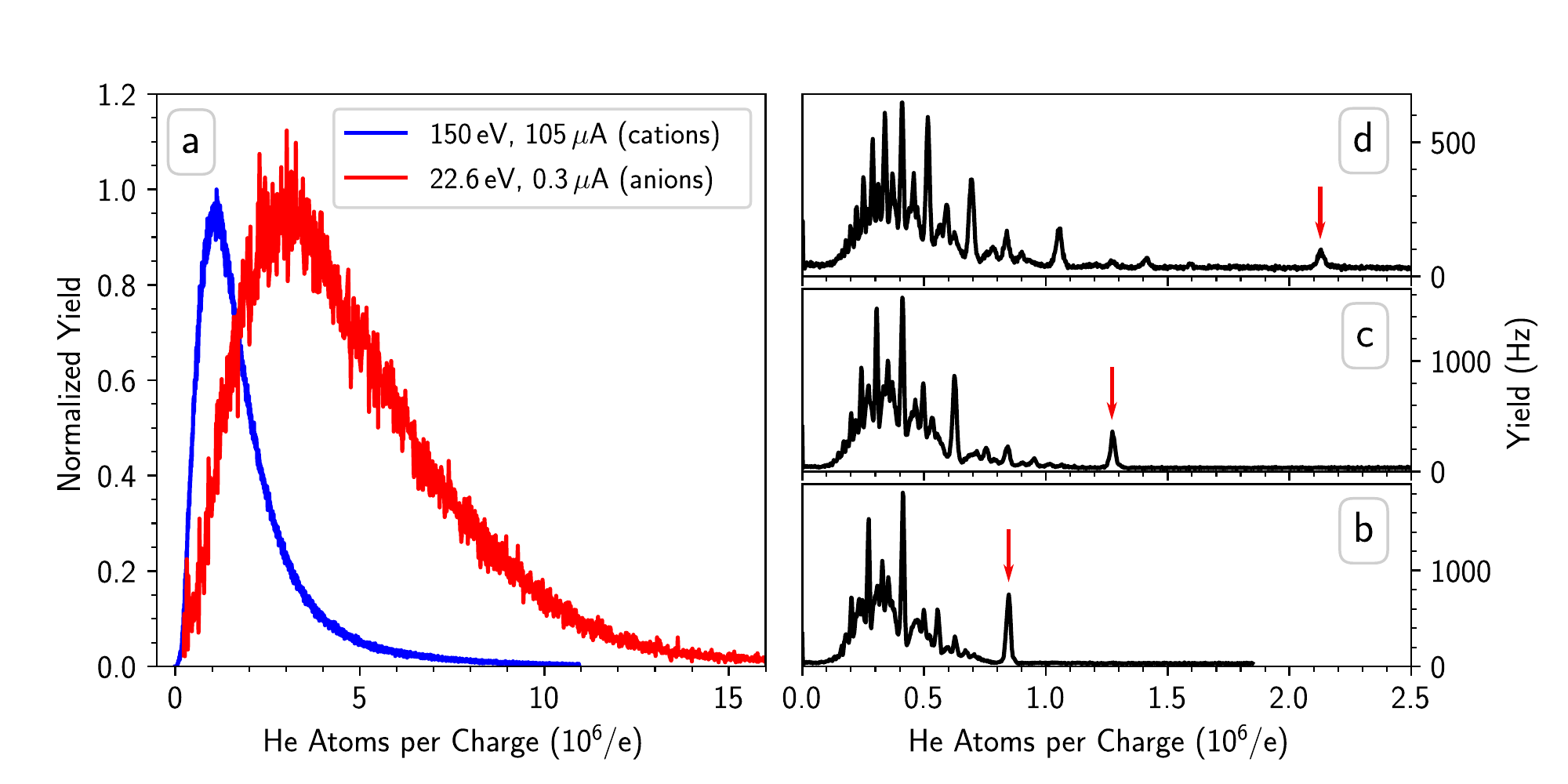} 
   \caption{{\textbf a} Mass per charge distributions of cationic He droplets (150\,eV electrons at 105.4\,$\mu$A, blue curve) and anionic droplets (22.6\,eV electrons at 0.3\,$\mu$A, red curve) by electron bombardment. The droplets were produced under identical conditions with a 8.5\,K nozzle temperature. The lower energy gives a distribution of essentially purely singly charged droplets that peaks near 3 million atoms and closely matches to the neutral distribution. The distribution of positively charged droplets is pushed to lower mass per charge ratios. {\textbf{b--d}} Distributions of He droplets that are $m/z$-selected slices from the blue distribution in panel a and ionized a second time. The parent ions have sizes of $8.5\times10^5$, $1.275\times10^6$, and $2.125\times10^6$ He atoms per charge (indicated by arrows), respectively, and the products all have rational fractions of the parent mass per charge ratio.} 
   \label{fig:1}
\end{figure*}

Droplets that are produced from the expansion of cooled and compressed He gas form broad log-normal size distributions in the size regime (millions of atoms) studied here \cite{Jiang:1992aa,Mauracher:2018aa}. In Figure \ref{fig:1}a we show two size-per-charge distributions of He droplets formed under identical conditions, but where the current and energy of electrons in the first ion source differ. Here, droplets were ionized by electrons with kinetic energies of 22.6\,eV and 150\,eV, respectively, with the latter also at a higher electron current. The red dataset shows a broad log-normal size distribution of negatively charged droplets that peaks near 3 million He atoms per unit charge. Since negative charge centers are heliophobic and form voids in the droplets, they are readily expelled from the droplets if multiple charges are present \cite{Farnik:1997bd,Mauracher:2018aa}. The distribution of anions can thus be assumed to mainly contain only singly charged droplets and represent the neutral size distribution. At the higher electron energy, multiple positive charges may be formed in the He droplets. This increases the energy deposited into the droplets, which could cause them to boil off He atoms and shrink in size. However, as will be discussed, the dominant mechanism is the accumulation of charges in the droplets that leads to a decrease in their mass per charge ratio and, if the charge density is high enough, to the ejection of low mass fragment ions. The blue dataset shows that this leads to an apparent size distribution that, while still close to log-normal in shape, now peaks at less than 1 million He atoms per unit charge.

 \begin{figure*}[] %  figure placement: here, top, bottom, or page
   \centering

   \includegraphics[width=5.5in]{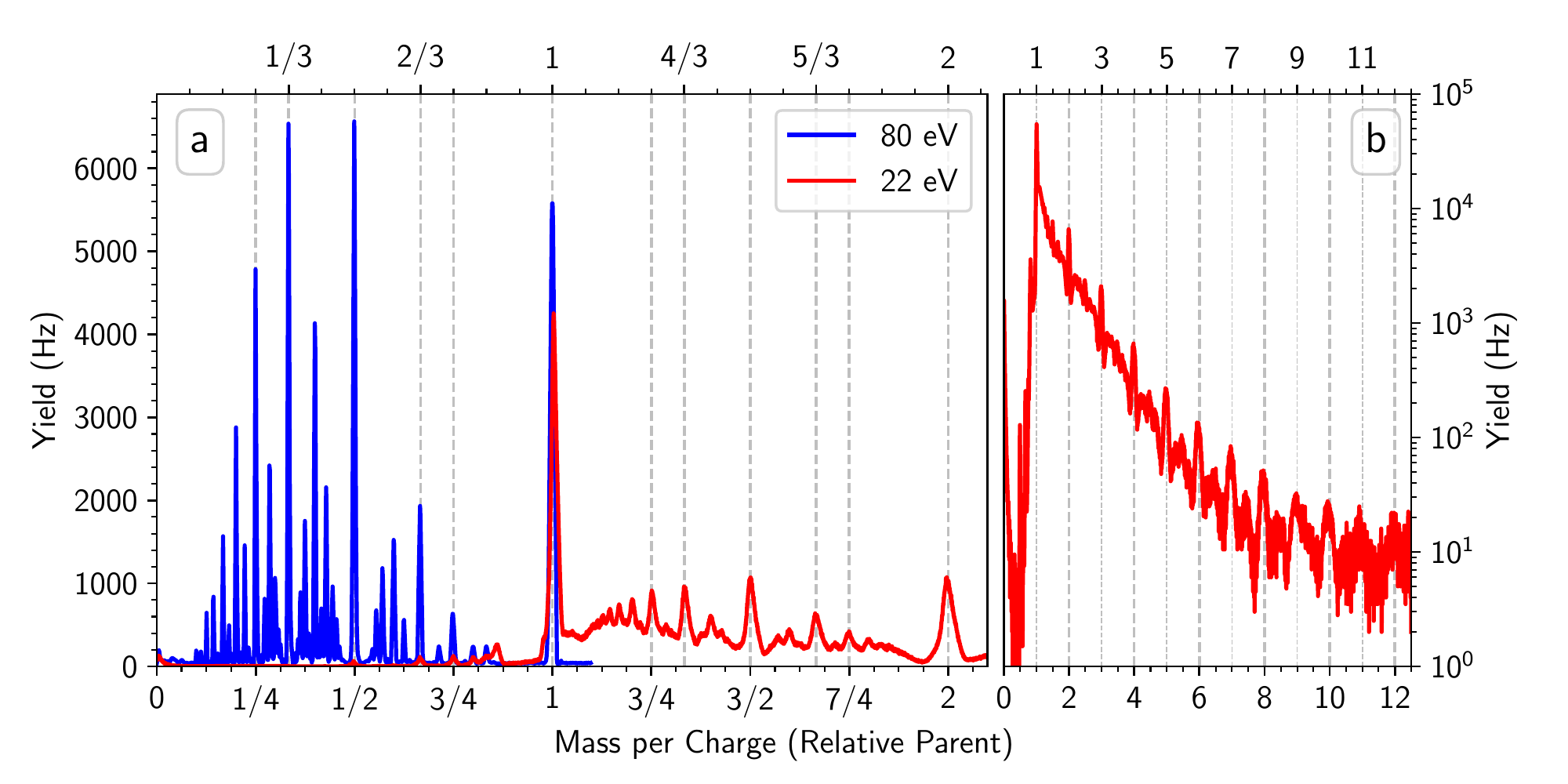} 

   \caption{\textbf{a} Mass-per-charge-selected ($3.8\times10^6$ He atoms per charge from 7\,K nozzle) droplets ionized a second time with 80\,eV (blue) and 22\,eV (red) electrons. The higher energy leads mainly to an increase in charge state and narrow peaks with mass per charge ratios at distinct rational fractions of the parent. At 22\,eV, the droplets are for the most part partially neutralized by helium anions, causing their mass per charge ratios to increase. \textbf{b} Wider range spectrum which shows that parent droplets containing up to 12 charges have had their net charge state reduced to +1. Peaks at half-integer positions show that droplets containing up to at least 17 charges have had their net charge reduced to +2. } 

   \label{fig:2}
\end{figure*} 

A novel feature of the experimental setup is that we can mass-select droplets after the first ionization source and let them interact with energetic electrons for a second time. In Figures \ref{fig:1}b, c, and d, we show some illustrative distributions of charged He droplets that have been $m/z$-selected, with narrow size distributions of about $8.5\times10^5$, $1.3\times10^6$, and $2.1\times10^6$ He atoms per charge, respectively. All three of these distributions were produced in the same way from the same initial distribution, with 150\,eV electrons in the first ionization source and 200\,eV electrons in the second (at 105.4\,$\mu$A and 197.1\,$\mu$A, respectively). Now, instead of the intact distributions being shifted continuously towards lower masses, we see a series of narrow peaks (FWHMs $\sim$3\% of mass per charge ratio, limited by experimental resolution) centered around rational fractions of the mass per charge ratio of the parent clusters. While one might expect that this effect is caused by the nearly symmetrical fission of large multiply charged droplets into smaller droplets with lower charge states, this is not what we are actually observing. Instead, we find that these peaks result from stable, multiply charged droplets. The fractional relative mass per charge ratios of the droplets correspond to the ratios between the charge state of the parent droplet and those of the daughter droplets, $z_p/z_d$, which remain intact after the second ionizing process. By tuning the settings of the two ion sources, as well as the mass per charge ratio of the parent droplets that are selected after the first ionization process, we can discern parent and daughter droplets with up to several tens of charges that give a range of different rational fractions of the parent mass per charge ratio. It is the wide range of higher charge states present in the parent droplets and the overlap of the numerous daughter droplets with different charge stats that are responsible for the broad features seen below the the narrow peaks. Interestingly, all three panels show a pileup of peaks around the same mass per charge ratio, about $3\times10^5$ He atoms per charge. This specific value depends on the experimental conditions, but the trend is easily reproducible in different measurements and is a result of the different electron impact cross sections of droplets in the sample.  

In Figure \ref{fig:2}a we show mass spectra from $m/z$-selected parent droplets containing $3.8\times10^6$ He atoms per charge (formed by 40\,eV electrons in the first ion source) that have been impacted a second time with electrons at kinetic energies of 22\,eV and 80\,eV, respectively. The different setting used compared to Figure \ref{fig:1} were chosen to best highlight the buildup of discrete charges in the selected droplets. At the higher energy, the impacting electrons may produce several He$^+$ ions (IE(He) = 24.6\,eV) along their trajectory through a droplet, resulting in an increase in the net positive charge. As the parent droplets carry several different charge states, all with the same mass per charge ratio, the result is a swarm of different daughter peaks. For example, the peak at $1/4$ is formed by the ions selected in the first mass selection stage having their net charge increase by a factor of $4$ by the subsequent ionization in the second stage. The close match between daughter peak position and rational fractions of the parent is remarkable and suggests that little evaporation of helium takes place in these secondary charging events. Some of the most prominent narrow peaks are visible at mass per charge ratios relative to the parents of $3/4$, $2/3$, $1/2$, $1/3$, and $1/4$, although several other peaks are clearly visible at other fractions.

When lower energy electrons are used to impact the $m/z$-selected He droplets it is possible to neutralize and reduce their charge states. This can be seen from the red curves in Figures \ref{fig:2}a and \ref{fig:2}b. Here, the energy of the electrons from the second ionization source has been tuned to below the ionization threshold of He. At this energy the electrons may lose energy as they scatter off of the neutral He, forming electronically excited He$^{*}$. The slowed electrons, which form voids in the droplets, or the He$^{*-}$ that are formed by the capture of the slow electrons by He$^*$\cite{Mauracher:2014ab}, may then neutralize positive charge centers in the parent droplets, effectively \emph{increasing} their mass per charge ratios. Numerous examples of multiply charged droplets having their net charge reduced are clearly seen in Figures \ref{fig:2}a and \ref{fig:2}b where several peaks with mass per charge ratios with rational numbers greater than one are visible. The peaks with integer values are predominantly from daughter droplets with a charge state of +1, originating from parents with up to 12 charges, all with the same initial mass per charge. Likewise, several half-integer peaks (up to at least 8.5 times the mass per charge ratio of the parent) are visible from even larger daughter droplets that still contain two positive charges.

\begin{figure}[h] %  figure placement: here, top, bottom, or page
   \centering
     %\begin{minipage}[b]{0.45\textwidth}

   \includegraphics[width=3in]{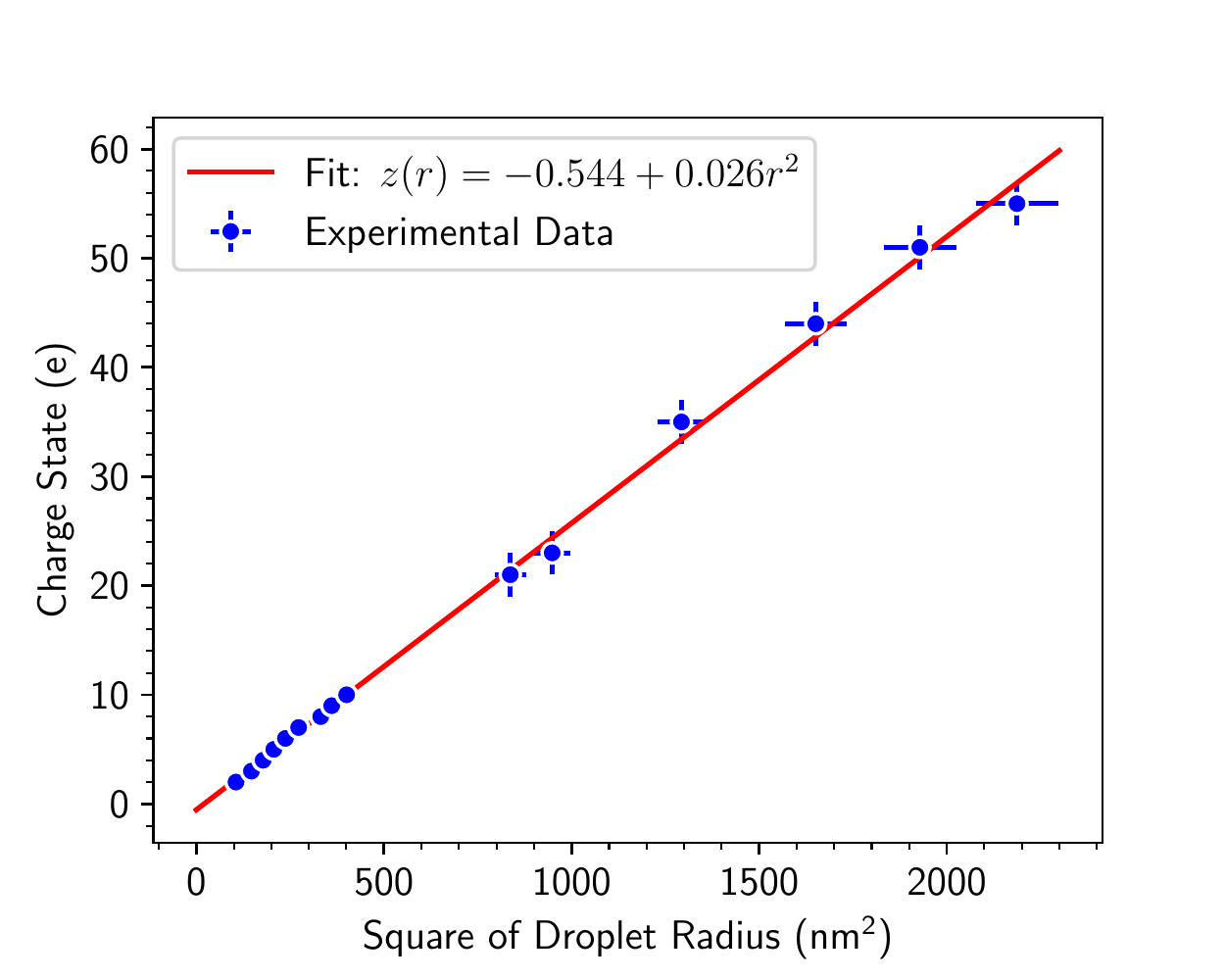} 
       % \end{minipage}
          %   \begin{minipage}[b]{0.45\textwidth}
   %\includegraphics[width=3.0in]{Pos_sizes.pdf} 
      %     \end{minipage}
   \caption{Plot of the droplet charge state versus the square of the minimum droplet radius needed to form that particular charge state. The red curve is a fit to the experimental data. The horizontal error bars originate from the statistical uncertainties in determining the sizes of the droplets and the vertical bars from the uncertainties in the appearance of specific charge numbers amongst the series of higher charge states.} 

   \label{fig:4}

\end{figure}

There is a general consensus that the charge centers of multiply charged He droplets are promptly ejected as small He$_n^+$ clusters \cite{Callicoatt:1998aa}, leaving at most a single charge in the remaining droplet. In the present measurements we resolve and identify multiply charged He droplets containing up several tens of positive charges, presumably in the form of solvated He$_n^+$ cores \cite{Callicoatt:1998aa,Ellis:2007aa,Mateo:2014aa}. In Figure \ref{fig:4} we show the appearance sizes of droplets for a range of high charge states. We find that the critical size of a droplet that can contain a given number of charges scales with the square of the radius of the droplet (determined by assuming spherical geometries and using the mean density of bulk superfluid He). This dependence could indicate that the appearance size of a multiply charged droplet scales with the cross section of the ionization processes. Another possibility is that the critical size is dictated by the multiple charges residing on the surface of the droplets, as would be expected for highly mobile interacting charge centers.

Doubly charged He droplets have been reported previously by F\'arn\'ik \emph{et al.}\ \cite{Farnik:1997bd}, who identified a threshold size of approximately $2\times10^{5}$ He atoms for observing these ions. In our measurements we find a significantly smaller appearance size. For the doubly charged droplets we measure a minimum size of $(1.00 \pm 0.05) \times 10^5$ atoms and our threshold for triply charged droplets is $(1.63 \pm 0.08) \times 10^5$. The largest systems we have measured the appearance sizes for are droplets containing 55 charges, the smallest of which consists of $(9.35\pm0.47)\times 10^{6}$ atoms. The reason for the discrepancy in appearance size between our measurements and those by F\'arn\'ik \emph{et al.}\ \cite{Farnik:1997bd} is unclear, but could be the result of limitations in the older experiments (e.g.\ fixed electron energy). Using a classical liquid droplet model, Echt \emph{et al.}\ determined the critical size of He droplets containing up to 4 positive charges to be about $2\times10^5$ atoms \cite{Echt:1988aa}. Within this framework, for droplets with continuous charge distributions, the critical sizes of higher charge states can be determined as $n(z) = z^2/(z_c^2/n_c)$, where $n_c$ is the known critical size of droplets with $z_c$ charges \cite{Saunders:1992aa}. With this, the predicted appearance size of droplets containing 55 charges is $3.8\times10^7$ He atoms, about four times larger than the experimentally measured limit. This discrepancy is consistent with results for multiply charged Ne droplets \cite{Mahr:2007aa} where it was explained by quantum effects and discrete charge distributions in the real droplets. The comparison with the model and previous results with charge rare gas droplets suggests that the cohesive forces in the He droplets are enough to explain the stability of our highly charged systems \cite{Mahr:2007aa}. Above the charge states shown in Figure \ref{fig:4}, larger droplets with even higher charge states are expected to remain stable but unresolved in our measurements as the mass per charge selected, singly charged parent droplets used will be too heavy to be deflected in our electrostatic sectors. Based on the source settings, the largest droplets we can produce are expected to have radii greater than 1\,$\mu$m ($>10^{10}$ He atoms), which could contain many thousands of charges. Noteworthy is that we find no evidence for ongoing droplet decay after the highly charged droplets are produced, indicating that they are indeed stable on the ms timescale of our experiment.

For a spherical droplet containing more than a few tens of thousands of atoms, charges produced by the electron impact will initially be situated near the surface facing the electron source \cite{Ellis:2007aa}. However, the charges will be highly mobile in the superfluid and should swiftly restructure to minimize the total repulsion energy. The positions of the charge centers in the stable, multiply charged droplets could therefore be considered to be similar to the solutions of the Thomson problem of point charges confined in a sphere \cite{Thomson:1904aa}, as has been shown for mobile charges in other liquids \cite{Consta:2018aa}. Droplets with an overabundance of charges appear to behave similar to classical liquids as they approach the Rayleigh stability limit \cite{Rayleigh:1882aa}, losing only small portions of mass as charges are expelled \cite{Duft:2002gh,Duft:2003id}. The expelled charged centers likely consist of small He$_n^+$ units in densely packed Atkins snowballs \cite{Atkins:1959aa}, ions that are commonly found in experiments limited to studying lower masses \cite{Mauracher:2018aa,Stephens:1983aa,Buchenau:1990aa}. Given the low interaction energy of He atoms, it is also possible that the charges lead to a shell of densely packed He$_n^+$ snowballs around the center of the droplets where the density is lowered, which could ultimately lead to an empty void forming in the center akin to a soap bubble.

In experiments where \emph{neutral} He droplets are doped with atomic or molecular species and then ionized, the small charged products that can be studied there appear to only constitute a fraction of the overall charge, since these new results show that a large number of the charges remain in the droplets. This opens the door to new experimental techniques where \emph{multiply charged} droplets are seeded with dopants. For example, charged droplets could be used as a weakly interacting matrix for ion spectroscopy where, a single droplet can provide multiple, separated ion nucleation sites. This approach has the potential to provide a new form of spectroscopic experiments facilitated by helium-tagging and promises high signal levels because of the multiple sites available in each droplet. Each charge center can also be used as a distinct nucleation site for the production of clusters and nanoparticles. Since the cross section of the multiply charged droplets can be selected before dopant pickup, the size distribution of particles grown in this way can be narrowed and more finely tuned compared to the case when neutral droplets of random sizes are used to capture gas phase building blocks and grow nanoparticles and nanowires. %Additionally, these findings set the stage for further studies on the properties of charged quantum fluid droplets.

%\section*{Acknowledgments}
This work was supported by the EU commission, EFRE K-Regio FAENOMENAL EFRE 2016-4, the Austrian Science Fund FWF (P31149 and W1259) and the Swedish Research Council (contract No.\ 2016-06625). FZ acknowledges support from Brazilian agency CNPq.

%\printbibliography
\bibliography{Library.bib}

\begin{thebibliography}{26}%
\makeatletter
\providecommand \@ifxundefined [1]{%
 \@ifx{#1\undefined}
}%
\providecommand \@ifnum [1]{%
 \ifnum #1\expandafter \@firstoftwo
 \else \expandafter \@secondoftwo
 \fi
}%
\providecommand \@ifx [1]{%
 \ifx #1\expandafter \@firstoftwo
 \else \expandafter \@secondoftwo
 \fi
}%
\providecommand \natexlab [1]{#1}%
\providecommand \enquote  [1]{``#1''}%
\providecommand \bibnamefont  [1]{#1}%
\providecommand \bibfnamefont [1]{#1}%
\providecommand \citenamefont [1]{#1}%
\providecommand \href@noop [0]{\@secondoftwo}%
\providecommand \href [0]{\begingroup \@sanitize@url \@href}%
\providecommand \@href[1]{\@@startlink{#1}\@@href}%
\providecommand \@@href[1]{\endgroup#1\@@endlink}%
\providecommand \@sanitize@url [0]{\catcode `\\12\catcode `\$12\catcode
  `\&12\catcode `\#12\catcode `\^12\catcode `\_12\catcode `\%12\relax}%
\providecommand \@@startlink[1]{}%
\providecommand \@@endlink[0]{}%
\providecommand \url  [0]{\begingroup\@sanitize@url \@url }%
\providecommand \@url [1]{\endgroup\@href {#1}{\urlprefix }}%
\providecommand \urlprefix  [0]{URL }%
\providecommand \Eprint [0]{\href }%
\providecommand \doibase [0]{http://dx.doi.org/}%
\providecommand \selectlanguage [0]{\@gobble}%
\providecommand \bibinfo  [0]{\@secondoftwo}%
\providecommand \bibfield  [0]{\@secondoftwo}%
\providecommand \translation [1]{[#1]}%
\providecommand \BibitemOpen [0]{}%
\providecommand \bibitemStop [0]{}%
\providecommand \bibitemNoStop [0]{.\EOS\space}%
\providecommand \EOS [0]{\spacefactor3000\relax}%
\providecommand \BibitemShut  [1]{\csname bibitem#1\endcsname}%
\let\auto@bib@innerbib\@empty
%</preamble>
\bibitem [{\citenamefont {Northby}(2001)}]{Northby:2001aa}%
  \BibitemOpen
  \bibfield  {author} {\bibinfo {author} {\bibfnamefont {J.~A.}\ \bibnamefont
  {Northby}},\ }\bibfield  {booktitle} {\emph {\bibinfo {booktitle} {The
  Journal of Chemical Physics}},\ }\href {\doibase 10.1063/1.1418249}
  {\bibfield  {journal} {\bibinfo  {journal} {The Journal of Chemical Physics}\
  }\textbf {\bibinfo {volume} {115}},\ \bibinfo {pages} {10065} (\bibinfo
  {year} {2001})}\BibitemShut {NoStop}%
\bibitem [{\citenamefont {Toennies}\ and\ \citenamefont
  {Vilesov}(2004)}]{Toennies:2004aa}%
  \BibitemOpen
  \bibfield  {author} {\bibinfo {author} {\bibfnamefont {J.~P.}\ \bibnamefont
  {Toennies}}\ and\ \bibinfo {author} {\bibfnamefont {A.~F.}\ \bibnamefont
  {Vilesov}},\ }\href {\doibase 10.1002/anie.200300611} {\bibfield  {journal}
  {\bibinfo  {journal} {Angewandte Chemie International Edition}\ }\textbf
  {\bibinfo {volume} {43}},\ \bibinfo {pages} {2622} (\bibinfo {year}
  {2004})}\BibitemShut {NoStop}%
\bibitem [{\citenamefont {Stienkemeier}\ and\ \citenamefont
  {Lehmann}(2006)}]{Stienkemeier:2006aa}%
  \BibitemOpen
  \bibfield  {author} {\bibinfo {author} {\bibfnamefont {F.}~\bibnamefont
  {Stienkemeier}}\ and\ \bibinfo {author} {\bibfnamefont {K.~K.}\ \bibnamefont
  {Lehmann}},\ }\href@noop {} {\bibfield  {journal} {\bibinfo  {journal}
  {Journal of Physics B: Atomic, Molecular and Optical Physics}\ }\textbf
  {\bibinfo {volume} {39}},\ \bibinfo {pages} {R127} (\bibinfo {year}
  {2006})}\BibitemShut {NoStop}%
\bibitem [{\citenamefont {Mauracher}\ \emph {et~al.}(2018)\citenamefont
  {Mauracher}, \citenamefont {Echt}, \citenamefont {Ellis}, \citenamefont
  {Yang}, \citenamefont {Bohme}, \citenamefont {Postler}, \citenamefont
  {Kaiser}, \citenamefont {Denifl},\ and\ \citenamefont
  {Scheier}}]{Mauracher:2018aa}%
  \BibitemOpen
  \bibfield  {author} {\bibinfo {author} {\bibfnamefont {A.}~\bibnamefont
  {Mauracher}}, \bibinfo {author} {\bibfnamefont {O.}~\bibnamefont {Echt}},
  \bibinfo {author} {\bibfnamefont {A.~M.}\ \bibnamefont {Ellis}}, \bibinfo
  {author} {\bibfnamefont {S.}~\bibnamefont {Yang}}, \bibinfo {author}
  {\bibfnamefont {D.~K.}\ \bibnamefont {Bohme}}, \bibinfo {author}
  {\bibfnamefont {J.}~\bibnamefont {Postler}}, \bibinfo {author} {\bibfnamefont
  {A.}~\bibnamefont {Kaiser}}, \bibinfo {author} {\bibfnamefont
  {S.}~\bibnamefont {Denifl}}, \ and\ \bibinfo {author} {\bibfnamefont
  {P.}~\bibnamefont {Scheier}},\ }\href@noop {} {\bibfield  {journal} {\bibinfo
   {journal} {Physics Reports}\ }\textbf {\bibinfo {volume} {751}},\ \bibinfo
  {pages} {1} (\bibinfo {year} {2018})}\BibitemShut {NoStop}%
\bibitem [{\citenamefont {Campbell}\ \emph {et~al.}(2015)\citenamefont
  {Campbell}, \citenamefont {Holz}, \citenamefont {Gerlich},\ and\
  \citenamefont {Maier}}]{Campbell:2015aa}%
  \BibitemOpen
  \bibfield  {author} {\bibinfo {author} {\bibfnamefont {E.~K.}\ \bibnamefont
  {Campbell}}, \bibinfo {author} {\bibfnamefont {M.}~\bibnamefont {Holz}},
  \bibinfo {author} {\bibfnamefont {D.}~\bibnamefont {Gerlich}}, \ and\
  \bibinfo {author} {\bibfnamefont {J.~P.}\ \bibnamefont {Maier}},\ }\href@noop
  {} {\bibfield  {journal} {\bibinfo  {journal} {Nature}\ }\textbf {\bibinfo
  {volume} {523}},\ \bibinfo {pages} {322} (\bibinfo {year}
  {2015})}\BibitemShut {NoStop}%
\bibitem [{\citenamefont {Kuhn}\ \emph {et~al.}(2016)\citenamefont {Kuhn},
  \citenamefont {Renzler}, \citenamefont {Postler}, \citenamefont {Ralser},
  \citenamefont {Spieler}, \citenamefont {Simpson}, \citenamefont {Linnartz},
  \citenamefont {Tielens}, \citenamefont {Cami}, \citenamefont {Mauracher},
  \citenamefont {Wang}, \citenamefont {Alcam{\'\i}}, \citenamefont
  {Mart{\'\i}n}, \citenamefont {Beyer}, \citenamefont {Wester}, \citenamefont
  {Lindinger},\ and\ \citenamefont {Scheier}}]{Kuhn:2016aa}%
  \BibitemOpen
  \bibfield  {author} {\bibinfo {author} {\bibfnamefont {M.}~\bibnamefont
  {Kuhn}}, \bibinfo {author} {\bibfnamefont {M.}~\bibnamefont {Renzler}},
  \bibinfo {author} {\bibfnamefont {J.}~\bibnamefont {Postler}}, \bibinfo
  {author} {\bibfnamefont {S.}~\bibnamefont {Ralser}}, \bibinfo {author}
  {\bibfnamefont {S.}~\bibnamefont {Spieler}}, \bibinfo {author} {\bibfnamefont
  {M.}~\bibnamefont {Simpson}}, \bibinfo {author} {\bibfnamefont
  {H.}~\bibnamefont {Linnartz}}, \bibinfo {author} {\bibfnamefont {A.~G.
  G.~M.}\ \bibnamefont {Tielens}}, \bibinfo {author} {\bibfnamefont
  {J.}~\bibnamefont {Cami}}, \bibinfo {author} {\bibfnamefont {A.}~\bibnamefont
  {Mauracher}}, \bibinfo {author} {\bibfnamefont {Y.}~\bibnamefont {Wang}},
  \bibinfo {author} {\bibfnamefont {M.}~\bibnamefont {Alcam{\'\i}}}, \bibinfo
  {author} {\bibfnamefont {F.}~\bibnamefont {Mart{\'\i}n}}, \bibinfo {author}
  {\bibfnamefont {M.~K.}\ \bibnamefont {Beyer}}, \bibinfo {author}
  {\bibfnamefont {R.}~\bibnamefont {Wester}}, \bibinfo {author} {\bibfnamefont
  {A.}~\bibnamefont {Lindinger}}, \ and\ \bibinfo {author} {\bibfnamefont
  {P.}~\bibnamefont {Scheier}},\ }\href@noop {} {\bibfield  {journal} {\bibinfo
   {journal} {Nature Communications}\ }\textbf {\bibinfo {volume} {7}},\
  \bibinfo {pages} {13550} (\bibinfo {year} {2016})}\BibitemShut {NoStop}%
\bibitem [{\citenamefont {Gomez}\ \emph {et~al.}(2014)\citenamefont {Gomez},
  \citenamefont {Ferguson}, \citenamefont {Cryan}, \citenamefont {Bacellar},
  \citenamefont {Tanyag}, \citenamefont {Jones}, \citenamefont {Schorb},
  \citenamefont {Anielski}, \citenamefont {Belkacem}, \citenamefont {Bernando},
  \citenamefont {Boll}, \citenamefont {Bozek}, \citenamefont {Carron},
  \citenamefont {Chen}, \citenamefont {Delmas}, \citenamefont {Englert},
  \citenamefont {Epp}, \citenamefont {Erk}, \citenamefont {Foucar},
  \citenamefont {Hartmann}, \citenamefont {Hexemer}, \citenamefont {Huth},
  \citenamefont {Kwok}, \citenamefont {Leone}, \citenamefont {Ma},
  \citenamefont {Maia}, \citenamefont {Malmerberg}, \citenamefont {Marchesini},
  \citenamefont {Neumark}, \citenamefont {Poon}, \citenamefont {Prell},
  \citenamefont {Rolles}, \citenamefont {Rudek}, \citenamefont {Rudenko},
  \citenamefont {Seifrid}, \citenamefont {Siefermann}, \citenamefont {Sturm},
  \citenamefont {Swiggers}, \citenamefont {Ullrich}, \citenamefont {Weise},
  \citenamefont {Zwart}, \citenamefont {Bostedt}, \citenamefont {Gessner},\
  and\ \citenamefont {Vilesov}}]{Gomez:2014aa}%
  \BibitemOpen
  \bibfield  {author} {\bibinfo {author} {\bibfnamefont {L.~F.}\ \bibnamefont
  {Gomez}}, \bibinfo {author} {\bibfnamefont {K.~R.}\ \bibnamefont {Ferguson}},
  \bibinfo {author} {\bibfnamefont {J.~P.}\ \bibnamefont {Cryan}}, \bibinfo
  {author} {\bibfnamefont {C.}~\bibnamefont {Bacellar}}, \bibinfo {author}
  {\bibfnamefont {R.~M.~P.}\ \bibnamefont {Tanyag}}, \bibinfo {author}
  {\bibfnamefont {C.}~\bibnamefont {Jones}}, \bibinfo {author} {\bibfnamefont
  {S.}~\bibnamefont {Schorb}}, \bibinfo {author} {\bibfnamefont
  {D.}~\bibnamefont {Anielski}}, \bibinfo {author} {\bibfnamefont
  {A.}~\bibnamefont {Belkacem}}, \bibinfo {author} {\bibfnamefont
  {C.}~\bibnamefont {Bernando}}, \bibinfo {author} {\bibfnamefont
  {R.}~\bibnamefont {Boll}}, \bibinfo {author} {\bibfnamefont {J.}~\bibnamefont
  {Bozek}}, \bibinfo {author} {\bibfnamefont {S.}~\bibnamefont {Carron}},
  \bibinfo {author} {\bibfnamefont {G.}~\bibnamefont {Chen}}, \bibinfo {author}
  {\bibfnamefont {T.}~\bibnamefont {Delmas}}, \bibinfo {author} {\bibfnamefont
  {L.}~\bibnamefont {Englert}}, \bibinfo {author} {\bibfnamefont {S.~W.}\
  \bibnamefont {Epp}}, \bibinfo {author} {\bibfnamefont {B.}~\bibnamefont
  {Erk}}, \bibinfo {author} {\bibfnamefont {L.}~\bibnamefont {Foucar}},
  \bibinfo {author} {\bibfnamefont {R.}~\bibnamefont {Hartmann}}, \bibinfo
  {author} {\bibfnamefont {A.}~\bibnamefont {Hexemer}}, \bibinfo {author}
  {\bibfnamefont {M.}~\bibnamefont {Huth}}, \bibinfo {author} {\bibfnamefont
  {J.}~\bibnamefont {Kwok}}, \bibinfo {author} {\bibfnamefont {S.~R.}\
  \bibnamefont {Leone}}, \bibinfo {author} {\bibfnamefont {J.~H.~S.}\
  \bibnamefont {Ma}}, \bibinfo {author} {\bibfnamefont {F.~R. N.~C.}\
  \bibnamefont {Maia}}, \bibinfo {author} {\bibfnamefont {E.}~\bibnamefont
  {Malmerberg}}, \bibinfo {author} {\bibfnamefont {S.}~\bibnamefont
  {Marchesini}}, \bibinfo {author} {\bibfnamefont {D.~M.}\ \bibnamefont
  {Neumark}}, \bibinfo {author} {\bibfnamefont {B.}~\bibnamefont {Poon}},
  \bibinfo {author} {\bibfnamefont {J.}~\bibnamefont {Prell}}, \bibinfo
  {author} {\bibfnamefont {D.}~\bibnamefont {Rolles}}, \bibinfo {author}
  {\bibfnamefont {B.}~\bibnamefont {Rudek}}, \bibinfo {author} {\bibfnamefont
  {A.}~\bibnamefont {Rudenko}}, \bibinfo {author} {\bibfnamefont
  {M.}~\bibnamefont {Seifrid}}, \bibinfo {author} {\bibfnamefont {K.~R.}\
  \bibnamefont {Siefermann}}, \bibinfo {author} {\bibfnamefont {F.~P.}\
  \bibnamefont {Sturm}}, \bibinfo {author} {\bibfnamefont {M.}~\bibnamefont
  {Swiggers}}, \bibinfo {author} {\bibfnamefont {J.}~\bibnamefont {Ullrich}},
  \bibinfo {author} {\bibfnamefont {F.}~\bibnamefont {Weise}}, \bibinfo
  {author} {\bibfnamefont {P.}~\bibnamefont {Zwart}}, \bibinfo {author}
  {\bibfnamefont {C.}~\bibnamefont {Bostedt}}, \bibinfo {author} {\bibfnamefont
  {O.}~\bibnamefont {Gessner}}, \ and\ \bibinfo {author} {\bibfnamefont
  {A.~F.}\ \bibnamefont {Vilesov}},\ }\href {\doibase 10.1126/science.1252395}
  {\bibfield  {journal} {\bibinfo  {journal} {Science}\ }\textbf {\bibinfo
  {volume} {345}},\ \bibinfo {pages} {906} (\bibinfo {year}
  {2014})}\BibitemShut {NoStop}%
\bibitem [{\citenamefont {F{\'a}rn{\'\i}k}\ \emph {et~al.}(1997)\citenamefont
  {F{\'a}rn{\'\i}k}, \citenamefont {Henne}, \citenamefont {Samelin},\ and\
  \citenamefont {Toennies}}]{Farnik:1997bd}%
  \BibitemOpen
  \bibfield  {author} {\bibinfo {author} {\bibfnamefont {M.}~\bibnamefont
  {F{\'a}rn{\'\i}k}}, \bibinfo {author} {\bibfnamefont {U.}~\bibnamefont
  {Henne}}, \bibinfo {author} {\bibfnamefont {B.}~\bibnamefont {Samelin}}, \
  and\ \bibinfo {author} {\bibfnamefont {J.~P.}\ \bibnamefont {Toennies}},\
  }\href@noop {} {\bibfield  {journal} {\bibinfo  {journal} {Zeitschrift
  f{\"u}r Physik D Atoms, Molecules and Clusters}\ }\textbf {\bibinfo {volume}
  {40}},\ \bibinfo {pages} {93} (\bibinfo {year} {1997})}\BibitemShut {NoStop}%
\bibitem [{\citenamefont {Henne}(1996)}]{Henne:1996aa}%
  \BibitemOpen
  \bibfield  {author} {\bibinfo {author} {\bibfnamefont {U.}~\bibnamefont
  {Henne}},\ }\emph {\bibinfo {title} {Untersuchung gro{\ss}er durch
  Elektronensto{\ss} erzeugter negativer und positiver Helium-Clusterionen}},\
  \href@noop {} {Ph.D. thesis},\ \bibinfo  {school} {Max-Panck-Institut F\"{u}r
  Str\"{o}mungsforschung} (\bibinfo {year} {1996})\BibitemShut {NoStop}%
\bibitem [{\citenamefont {Henne}\ and\ \citenamefont
  {Toennies}(1998)}]{Henne:1998aa}%
  \BibitemOpen
  \bibfield  {author} {\bibinfo {author} {\bibfnamefont {U.}~\bibnamefont
  {Henne}}\ and\ \bibinfo {author} {\bibfnamefont {J.~P.}\ \bibnamefont
  {Toennies}},\ }\bibfield  {booktitle} {\emph {\bibinfo {booktitle} {The
  Journal of Chemical Physics}},\ }\href {\doibase 10.1063/1.476385} {\bibfield
   {journal} {\bibinfo  {journal} {The Journal of Chemical Physics}\ }\textbf
  {\bibinfo {volume} {108}},\ \bibinfo {pages} {9327} (\bibinfo {year}
  {1998})}\BibitemShut {NoStop}%
\bibitem [{\citenamefont {Jiang}\ and\ \citenamefont
  {Northby}(1992)}]{Jiang:1992aa}%
  \BibitemOpen
  \bibfield  {author} {\bibinfo {author} {\bibfnamefont {T.}~\bibnamefont
  {Jiang}}\ and\ \bibinfo {author} {\bibfnamefont {J.~A.}\ \bibnamefont
  {Northby}},\ }\href {\doibase 10.1103/PhysRevLett.68.2620} {\bibfield
  {journal} {\bibinfo  {journal} {Phys. Rev. Lett.}\ }\textbf {\bibinfo
  {volume} {68}},\ \bibinfo {pages} {2620} (\bibinfo {year}
  {1992})}\BibitemShut {NoStop}%
\bibitem [{\citenamefont {Mauracher}\ \emph {et~al.}(2014)\citenamefont
  {Mauracher}, \citenamefont {Daxner}, \citenamefont {Postler}, \citenamefont
  {Huber}, \citenamefont {Denifl}, \citenamefont {Scheier},\ and\ \citenamefont
  {Toennies}}]{Mauracher:2014ab}%
  \BibitemOpen
  \bibfield  {author} {\bibinfo {author} {\bibfnamefont {A.}~\bibnamefont
  {Mauracher}}, \bibinfo {author} {\bibfnamefont {M.}~\bibnamefont {Daxner}},
  \bibinfo {author} {\bibfnamefont {J.}~\bibnamefont {Postler}}, \bibinfo
  {author} {\bibfnamefont {S.~E.}\ \bibnamefont {Huber}}, \bibinfo {author}
  {\bibfnamefont {S.}~\bibnamefont {Denifl}}, \bibinfo {author} {\bibfnamefont
  {P.}~\bibnamefont {Scheier}}, \ and\ \bibinfo {author} {\bibfnamefont
  {J.~P.}\ \bibnamefont {Toennies}},\ }\bibfield  {booktitle} {\emph {\bibinfo
  {booktitle} {The Journal of Physical Chemistry Letters}},\ }\href {\doibase
  10.1021/jz500917z} {\bibfield  {journal} {\bibinfo  {journal} {The Journal of
  Physical Chemistry Letters}\ }\textbf {\bibinfo {volume} {5}},\ \bibinfo
  {pages} {2444} (\bibinfo {year} {2014})}\BibitemShut {NoStop}%
\bibitem [{\citenamefont {Callicoatt}\ \emph {et~al.}(1998)\citenamefont
  {Callicoatt}, \citenamefont {F{\"o}rde}, \citenamefont {Jung}, \citenamefont
  {Ruchti},\ and\ \citenamefont {Janda}}]{Callicoatt:1998aa}%
  \BibitemOpen
  \bibfield  {author} {\bibinfo {author} {\bibfnamefont {B.~E.}\ \bibnamefont
  {Callicoatt}}, \bibinfo {author} {\bibfnamefont {K.}~\bibnamefont
  {F{\"o}rde}}, \bibinfo {author} {\bibfnamefont {L.~F.}\ \bibnamefont {Jung}},
  \bibinfo {author} {\bibfnamefont {T.}~\bibnamefont {Ruchti}}, \ and\ \bibinfo
  {author} {\bibfnamefont {K.~C.}\ \bibnamefont {Janda}},\ }\bibfield
  {booktitle} {\emph {\bibinfo {booktitle} {The Journal of Chemical Physics}},\
  }\href {\doibase 10.1063/1.477713} {\bibfield  {journal} {\bibinfo  {journal}
  {The Journal of Chemical Physics}\ }\textbf {\bibinfo {volume} {109}},\
  \bibinfo {pages} {10195} (\bibinfo {year} {1998})}\BibitemShut {NoStop}%
\bibitem [{\citenamefont {Ellis}\ and\ \citenamefont
  {Yang}(2007)}]{Ellis:2007aa}%
  \BibitemOpen
  \bibfield  {author} {\bibinfo {author} {\bibfnamefont {A.~M.}\ \bibnamefont
  {Ellis}}\ and\ \bibinfo {author} {\bibfnamefont {S.}~\bibnamefont {Yang}},\
  }\href@noop {} {\bibfield  {journal} {\bibinfo  {journal} {Phys. Rev. A}\
  }\textbf {\bibinfo {volume} {76}},\ \bibinfo {pages} {032714} (\bibinfo
  {year} {2007})}\BibitemShut {NoStop}%
\bibitem [{\citenamefont {Mateo}\ and\ \citenamefont
  {Eloranta}(2014)}]{Mateo:2014aa}%
  \BibitemOpen
  \bibfield  {author} {\bibinfo {author} {\bibfnamefont {D.}~\bibnamefont
  {Mateo}}\ and\ \bibinfo {author} {\bibfnamefont {J.}~\bibnamefont
  {Eloranta}},\ }\bibfield  {booktitle} {\emph {\bibinfo {booktitle} {The
  Journal of Physical Chemistry A}},\ }\href {\doibase 10.1021/jp501451y}
  {\bibfield  {journal} {\bibinfo  {journal} {The Journal of Physical Chemistry
  A}\ }\textbf {\bibinfo {volume} {118}},\ \bibinfo {pages} {6407} (\bibinfo
  {year} {2014})}\BibitemShut {NoStop}%
\bibitem [{\citenamefont {Echt}\ \emph {et~al.}(1988)\citenamefont {Echt},
  \citenamefont {Kreisle}, \citenamefont {Recknagel}, \citenamefont {Saenz},
  \citenamefont {Casero},\ and\ \citenamefont {Soler}}]{Echt:1988aa}%
  \BibitemOpen
  \bibfield  {author} {\bibinfo {author} {\bibfnamefont {O.}~\bibnamefont
  {Echt}}, \bibinfo {author} {\bibfnamefont {D.}~\bibnamefont {Kreisle}},
  \bibinfo {author} {\bibfnamefont {E.}~\bibnamefont {Recknagel}}, \bibinfo
  {author} {\bibfnamefont {J.~J.}\ \bibnamefont {Saenz}}, \bibinfo {author}
  {\bibfnamefont {R.}~\bibnamefont {Casero}}, \ and\ \bibinfo {author}
  {\bibfnamefont {J.~M.}\ \bibnamefont {Soler}},\ }\href {\doibase
  10.1103/PhysRevA.38.3236} {\bibfield  {journal} {\bibinfo  {journal} {Phys.
  Rev. A}\ }\textbf {\bibinfo {volume} {38}},\ \bibinfo {pages} {3236}
  (\bibinfo {year} {1988})}\BibitemShut {NoStop}%
\bibitem [{\citenamefont {Saunders}(1992)}]{Saunders:1992aa}%
  \BibitemOpen
  \bibfield  {author} {\bibinfo {author} {\bibfnamefont {W.~A.}\ \bibnamefont
  {Saunders}},\ }\href {\doibase 10.1103/PhysRevA.46.7028} {\bibfield
  {journal} {\bibinfo  {journal} {Phys. Rev. A}\ }\textbf {\bibinfo {volume}
  {46}},\ \bibinfo {pages} {7028} (\bibinfo {year} {1992})}\BibitemShut
  {NoStop}%
\bibitem [{\citenamefont {M\"ahr}\ \emph {et~al.}(2007)\citenamefont {M\"ahr},
  \citenamefont {Zappa}, \citenamefont {Denifl}, \citenamefont {Kubala},
  \citenamefont {Echt}, \citenamefont {M\"ark},\ and\ \citenamefont
  {Scheier}}]{Mahr:2007aa}%
  \BibitemOpen
  \bibfield  {author} {\bibinfo {author} {\bibfnamefont {I.}~\bibnamefont
  {M\"ahr}}, \bibinfo {author} {\bibfnamefont {F.}~\bibnamefont {Zappa}},
  \bibinfo {author} {\bibfnamefont {S.}~\bibnamefont {Denifl}}, \bibinfo
  {author} {\bibfnamefont {D.}~\bibnamefont {Kubala}}, \bibinfo {author}
  {\bibfnamefont {O.}~\bibnamefont {Echt}}, \bibinfo {author} {\bibfnamefont
  {T.~D.}\ \bibnamefont {M\"ark}}, \ and\ \bibinfo {author} {\bibfnamefont
  {P.}~\bibnamefont {Scheier}},\ }\href {\doibase
  10.1103/PhysRevLett.98.023401} {\bibfield  {journal} {\bibinfo  {journal}
  {Phys. Rev. Lett.}\ }\textbf {\bibinfo {volume} {98}},\ \bibinfo {pages}
  {023401} (\bibinfo {year} {2007})}\BibitemShut {NoStop}%
\bibitem [{\citenamefont {Thomson}(1904)}]{Thomson:1904aa}%
  \BibitemOpen
  \bibfield  {author} {\bibinfo {author} {\bibfnamefont {J.~J.}\ \bibnamefont
  {Thomson}},\ }\bibfield  {booktitle} {\emph {\bibinfo {booktitle} {The
  London, Edinburgh, and Dublin Philosophical Magazine and Journal of
  Science}},\ }\href {\doibase 10.1080/14786440409463107} {\bibfield  {journal}
  {\bibinfo  {journal} {The London, Edinburgh, and Dublin Philosophical
  Magazine and Journal of Science}\ }\textbf {\bibinfo {volume} {7}},\ \bibinfo
  {pages} {237} (\bibinfo {year} {1904})}\BibitemShut {NoStop}%
\bibitem [{\citenamefont {Consta}\ \emph {et~al.}(2018)\citenamefont {Consta},
  \citenamefont {Oh}, \citenamefont {Sharawy},\ and\ \citenamefont
  {Malevanets}}]{Consta:2018aa}%
  \BibitemOpen
  \bibfield  {author} {\bibinfo {author} {\bibfnamefont {S.}~\bibnamefont
  {Consta}}, \bibinfo {author} {\bibfnamefont {M.~I.}\ \bibnamefont {Oh}},
  \bibinfo {author} {\bibfnamefont {M.}~\bibnamefont {Sharawy}}, \ and\
  \bibinfo {author} {\bibfnamefont {A.}~\bibnamefont {Malevanets}},\ }\bibfield
   {booktitle} {\emph {\bibinfo {booktitle} {The Journal of Physical Chemistry
  A}},\ }\href {\doibase 10.1021/acs.jpca.8b01404} {\bibfield  {journal}
  {\bibinfo  {journal} {The Journal of Physical Chemistry A}\ }\textbf
  {\bibinfo {volume} {122}},\ \bibinfo {pages} {5239} (\bibinfo {year}
  {2018})}\BibitemShut {NoStop}%
\bibitem [{\citenamefont {Rayleigh}(1882)}]{Rayleigh:1882aa}%
  \BibitemOpen
  \bibfield  {author} {\bibinfo {author} {\bibfnamefont {L.}~\bibnamefont
  {Rayleigh}},\ }\bibfield  {booktitle} {\emph {\bibinfo {booktitle} {The
  London, Edinburgh, and Dublin Philosophical Magazine and Journal of
  Science}},\ }\href {\doibase 10.1080/14786448208628425} {\bibfield  {journal}
  {\bibinfo  {journal} {The London, Edinburgh, and Dublin Philosophical
  Magazine and Journal of Science}\ }\textbf {\bibinfo {volume} {14}},\
  \bibinfo {pages} {184} (\bibinfo {year} {1882})}\BibitemShut {NoStop}%
\bibitem [{\citenamefont {Duft}\ \emph {et~al.}(2002)\citenamefont {Duft},
  \citenamefont {Lebius}, \citenamefont {Huber}, \citenamefont {Guet},\ and\
  \citenamefont {Leisner}}]{Duft:2002gh}%
  \BibitemOpen
  \bibfield  {author} {\bibinfo {author} {\bibfnamefont {D.}~\bibnamefont
  {Duft}}, \bibinfo {author} {\bibfnamefont {H.}~\bibnamefont {Lebius}},
  \bibinfo {author} {\bibfnamefont {B.~A.}\ \bibnamefont {Huber}}, \bibinfo
  {author} {\bibfnamefont {C.}~\bibnamefont {Guet}}, \ and\ \bibinfo {author}
  {\bibfnamefont {T.}~\bibnamefont {Leisner}},\ }\href@noop {} {\bibfield
  {journal} {\bibinfo  {journal} {Phys. Rev. Lett.}\ }\textbf {\bibinfo
  {volume} {89}},\ \bibinfo {pages} {084503} (\bibinfo {year}
  {2002})}\BibitemShut {NoStop}%
\bibitem [{\citenamefont {Duft}\ \emph {et~al.}(2003)\citenamefont {Duft},
  \citenamefont {Achtzehn}, \citenamefont {M{\"u}ller}, \citenamefont {Huber},\
  and\ \citenamefont {Leisner}}]{Duft:2003id}%
  \BibitemOpen
  \bibfield  {author} {\bibinfo {author} {\bibfnamefont {D.}~\bibnamefont
  {Duft}}, \bibinfo {author} {\bibfnamefont {T.}~\bibnamefont {Achtzehn}},
  \bibinfo {author} {\bibfnamefont {R.}~\bibnamefont {M{\"u}ller}}, \bibinfo
  {author} {\bibfnamefont {B.~A.}\ \bibnamefont {Huber}}, \ and\ \bibinfo
  {author} {\bibfnamefont {T.}~\bibnamefont {Leisner}},\ }\href@noop {}
  {\bibfield  {journal} {\bibinfo  {journal} {Nature}\ }\textbf {\bibinfo
  {volume} {421}},\ \bibinfo {pages} {128} (\bibinfo {year}
  {2003})}\BibitemShut {NoStop}%
\bibitem [{\citenamefont {Atkins}(1959)}]{Atkins:1959aa}%
  \BibitemOpen
  \bibfield  {author} {\bibinfo {author} {\bibfnamefont {K.~R.}\ \bibnamefont
  {Atkins}},\ }\href {\doibase 10.1103/PhysRev.116.1339} {\bibfield  {journal}
  {\bibinfo  {journal} {Phys. Rev.}\ }\textbf {\bibinfo {volume} {116}},\
  \bibinfo {pages} {1339} (\bibinfo {year} {1959})}\BibitemShut {NoStop}%
\bibitem [{\citenamefont {Stephens}\ and\ \citenamefont
  {King}(1983)}]{Stephens:1983aa}%
  \BibitemOpen
  \bibfield  {author} {\bibinfo {author} {\bibfnamefont {P.~W.}\ \bibnamefont
  {Stephens}}\ and\ \bibinfo {author} {\bibfnamefont {J.~G.}\ \bibnamefont
  {King}},\ }\href {\doibase 10.1103/PhysRevLett.51.1538} {\bibfield  {journal}
  {\bibinfo  {journal} {Physical Review Letters}\ }\textbf {\bibinfo {volume}
  {51}},\ \bibinfo {pages} {1538} (\bibinfo {year} {1983})}\BibitemShut
  {NoStop}%
\bibitem [{\citenamefont {Buchenau}\ \emph {et~al.}(1990)\citenamefont
  {Buchenau}, \citenamefont {Knuth}, \citenamefont {Northby}, \citenamefont
  {Toennies},\ and\ \citenamefont {Winkler}}]{Buchenau:1990aa}%
  \BibitemOpen
  \bibfield  {author} {\bibinfo {author} {\bibfnamefont {H.}~\bibnamefont
  {Buchenau}}, \bibinfo {author} {\bibfnamefont {E.~L.}\ \bibnamefont {Knuth}},
  \bibinfo {author} {\bibfnamefont {J.}~\bibnamefont {Northby}}, \bibinfo
  {author} {\bibfnamefont {J.~P.}\ \bibnamefont {Toennies}}, \ and\ \bibinfo
  {author} {\bibfnamefont {C.}~\bibnamefont {Winkler}},\ }\href {\doibase
  10.1063/1.458275} {\bibfield  {journal} {\bibinfo  {journal} {The Journal of
  Chemical Physics}\ }\textbf {\bibinfo {volume} {92}},\ \bibinfo {pages}
  {6875} (\bibinfo {year} {1990})}\BibitemShut {NoStop}%
\end{thebibliography}%

\end{document}